\documentclass[aps,prl,preprint,superscriptaddress,amsmath,amssymb]{revtex4-1}
 \usepackage{graphicx}
 \usepackage{subfigure}
 \usepackage{dcolumn}
 \usepackage{bm}
 \usepackage{sidecap} 
 \usepackage{braket}

\begin{document}

\title{A frequency-stabilized source of single photons from a solid-state qubit} 

\author{Jonathan H. Prechtel}
\email[]{jonathan.prechtel@unibas.ch}
\homepage[]{http://nano-photonics.unibas.ch/}

\author{Andreas V. Kuhlmann}
\affiliation{Department of Physics, University of Basel, Klingelbergstrasse 82, CH-4056 Basel, Switzerland}

\author{Julien Houel}
\affiliation{Department of Physics, University of Basel, Klingelbergstrasse 82, CH-4056 Basel, Switzerland}
\affiliation{B\^{a}timent Alfred Kastler, Universit\'e Claude Bernad, F-69622 Lyon Villeurbanne,  France}

\author{Lukas Greuter}
\affiliation{Department of Physics, University of Basel, Klingelbergstrasse 82, CH-4056 Basel, Switzerland}

\author{Arne Ludwig}
\affiliation{Lehrstuhl f\"{u}r Angewandte Festk\"{o}rperphysik, Ruhr-Universit\"{a}t Bochum, Universit\"{a}tsstr. 150, D-44780 Bochum, Germany }
\author{Dirk Reuter}
\affiliation{Lehrstuhl f\"{u}r Angewandte Festk\"{o}rperphysik, Ruhr-Universit\"{a}t Bochum, Universit\"{a}tsstr. 150, D-44780 Bochum, Germany }
\affiliation{Department Physik, Universit\"{a}t Paderborn, Warburgerstr. 100, D-33098 Paderborn, Germany }
\author{Andreas D. Wieck}
\affiliation{Lehrstuhl f\"{u}r Angewandte Festk\"{o}rperphysik, Ruhr-Universit\"{a}t Bochum, Universit\"{a}tsstr. 150, D-44780 Bochum, Germany }

\author{Richard J. Warburton}
\affiliation{Department of Physics, University of Basel, Klingelbergstrasse 82, CH-4056 Basel, Switzerland}

\date{\today}

\begin{abstract}
Single quantum dots are solid-state emitters which mimic two-level atoms but with a highly enhanced spontaneous emission rate. A single quantum dot is the basis for a potentially excellent single photon source. One outstanding problem is that there is considerable noise in the emission frequency, making it very difficult to couple the quantum dot to another quantum system. We solve this problem here with a dynamic feedback technique that locks the quantum dot emission frequency to a reference. The incoherent scattering (resonance fluorescence) represents the single photon output whereas the coherent scattering (Rayleigh scattering) is used for the feedback control. The fluctuations in emission frequency are reduced to 20 MHz, just $\sim 5$\% of the quantum dot optical linewidth, even over several hours. By eliminating the $1/f$-like noise, the relative fluctuations in resonance fluorescence intensity are reduced to $\sim 10^{-5}$ at low frequency. Under these conditions, the antibunching dip in the resonance fluorescence is described extremely well by the two-level atom result. The technique represents a way of removing charge noise from a quantum device.
\end{abstract}


\maketitle 

\label{sec:Introduction}
Single photons are ideal carriers of quantum information. A quantum state stored in one of the degrees of freedom of the photon's wave packet (polarization, phase or time-bin) can be maintained over long distances. Single photons are therefore important in quantum communication,  for coupling remote stationary qubits, the basis of a quantum repeater, or for coupling different elements in a quantum device. Furthermore, single photons are the seed for a variety of quantum optics experiments. 

A key challenge is to develop a single photon source \cite{Lounis2005}. Key parameters are fidelity of the antibunching, flux, wavelength and photon indistinguishability. Remarkably, solid-state emitters are presently better able to meet these demands than atomic systems (single atoms or parametric down conversion). In particular, spontaneous emission from individual quantum dots embedded in an inorganic semiconductor is a very promising source of highly antibunched, high flux, indistinguishable photons \cite{Michler2000,Santori2002,Shields2007}. The antibunching, particularly with resonant excitation, is very high \cite{Muller2007}. The radiative lifetime is very short, typically just less than 1 ns \cite{Dalgarno2008a}. The flux is usually limited by the poor collection efficiency: most of the light is internally reflected at the GaAs-vacuum interface. However, this problem can be solved by nano-structuring the photonic modes to create a micro-cavity \cite{Vahala2003} or a photonic nanowire \cite{Claudon2010}. In the latter case, collection efficiencies of $\sim 70$\% have been achieved. The photon indistinguishability is very high for successive photons \cite{Santori2002}. Based on the optical linewidth, typically a factor of two above the transform limit when measured with resonant excitation \cite{Hogele2004,Atature2006,Houel2012,andi}, the indistinguishability is also reasonably high for photons emitted widely separated in time. Furthermore, a single quantum dot has also been developed as a spin qubit \cite{Warburton2013}, facilitating an interface between stationary qubits and photons \cite{Yilmaz2010,DeGreve2012,Gao2012}. 

Unlike a real atom, the exact transition wavelength of a quantum dot is not locked to any particular wavelength and varies considerably from quantum dot to quantum dot. However, the host semiconductor can be designed so that considerable possibilities for tuning the emission wavelength exist. Electric field tuning \cite{Warburton2002,Bennett2010} and strain tuning \cite{Seidl2006a,Joens2011} allow the emission wavelength to be tuned over several nanometres. A major problem remains.  The emission wavelength is not constant: it varies randomly over time, even in very controlled environments at low temperature. The culprit at low frequency is electrical noise in the semiconductor which shifts the emission wavelength via the Stark effect \cite{andi}. This noise has a $1/f$-like power spectrum resulting in, first, large and uncontrolled drifts at low frequencies and second, an undefined mean value. This noise, while poorly understood, is ubiquitous in semiconductors and makes it very difficult to couple an individual quantum dot to another quantum system, another quantum dot for instance, or an ensemble of cold atoms. We present here a new scheme which solves this problem: we create a stream of single photons with a wavelength which remains constant even over several hours. 

The output of our quantum device is a stream of single photons generated by resonance fluorescence (RF) from a single quantum dot. RF has considerable advantages over non-resonant excitation of photoluminescence: the linewidth is much lower\cite{Houel2012,andi} and the antibunching is much better. We lock the wavelength of the quantum device to a stable reference. We generate an error signal, a signal with large slope at its zero-crossing, by measuring the differential transmission, $\Delta T/T$,  simultaneously \cite{Alen2003,Hogele2004,Karrai2003}. The control variable is the voltage $V_g$ applied to a surface gate which influences the quantum dot frequency via the Stark effect. The performance of the feedback scheme is characterized by, first, measuring a series of snap-shots of the optical resonance to assess the residual frequency jitter; and second, by carrying out a full analysis of the noise in the RF. This scheme goes well beyond previous attempts at single emitter stabilization in the solid-state \cite{Acosta:2012,Akopian:2013}. The absolute frequency of the quantum dot emission is locked with an uncertainty of just $20$ MHz. We observe a reduction in the noise power up to a frequency of $\sim 100$ Hz, high enough to eliminate the substantial drifts at low frequency. Arguably, these low frequency fluctuations have a classical nature, reflecting charge noise in our solid-state device on millisecond or second time-scales. On much shorter time scales, there are clear quantum effects: the intensity correlation coefficient exhibits a clear dip between 0 and $\pm 2$ ns; electron spins have decoherence times in the $\mu$s regime \cite{Warburton2013}. As such, this experiment represents a first step towards bridging these time scales, i.e.\ quantum control of a solid-state emitter.
\begin{figure}
\centering
\includegraphics[width=\linewidth]{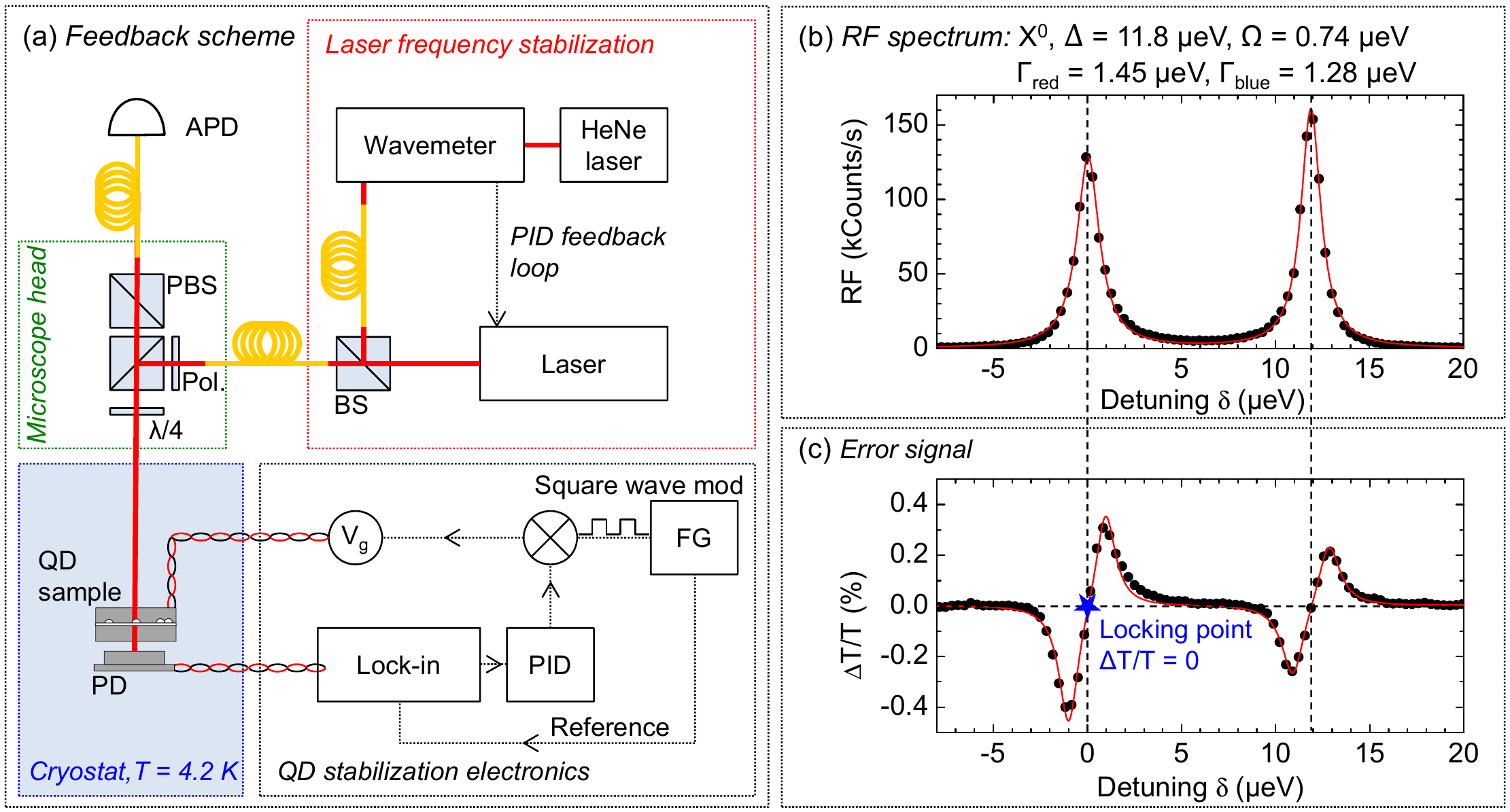}
\vspace*{-0.5cm}
\caption{
(a) Schematic view of the experiment. The narrowband laser is stabilized to a fixed frequency by a wavemeter which in turn is stabilized to a HeNe laser. Laser light is guided through optical fibers (yellow) and microscope optics before it is focused onto the sample, driving the X$^{0}$ transition resonantly (BS = beam-splitter, PBS = polarizing BS, Pol. = linear polarizer). Two simultaneous measurements of X$^{0}$ scattering are performed: resonance fluorescence (RF) with a dark field technique, detected with an avalanche photo diode (APD) and absorption with a photodiode (PD) underneath the sample. The dynamic stabilization is realized with an active PID feedback loop which corrects for fluctuations in the transition energy using the gate voltage $V_{g}$ and the square wave modulation of a function generator (FG). (b) RF signal of the fine-structure splitted X$^{0}$ emission of a single quantum dot at wavelength 936.5 nm, a power corresponding to a Rabi energy $\Omega$ of 0.74 $\mu$eV and a temperature of \mbox{4.2 K}. A detuning is achieved by sweeping the gate voltage. The solid red line is a Lorentzian fit to the data with linewidth $\Gamma$ = 1.28 $\mu$eV (309 MHz) and $\Gamma$ = 1.45 $\mu$eV (350 MHz) and with a fine structure splitting \mbox{$\Delta$ = 11.6 $\mu$eV}.
(c) The differential transmission ($\Delta T/T$) signal on the same quantum dot with integration time 100 ms per point. The red curve is a fit to the derivative of the two Lorentzians. The signal around the zero crossing point ($\Delta T/T = 0$) is used to generate an error signal for the feedback scheme.}
\label{fig:1}
\end{figure}
A sketch of the experimental concept is shown in \mbox{Fig.\ \ref{fig:1}(a)}. A linearly-polarized resonant laser is focused onto the sample surface and drives the optical transition. The resonance fluorescence of the quantum dot is collected with a polarization-based dark field technique \cite{Houel2012,Matthiesen2012,Yilmaz2010}, described in detail elsewhere \cite{andiRSI}. Simultaneously, the optical resonance is detected in transmission by superimposing a sub-linewidth modulation to the gate. The transmission signal arises from an interference of quantum dot scattering with the driving laser \cite{Karrai2003}. The incoherent part, i.e.\ the resonance fluorescence, averages to zero in transmission; what is detected instead is the coherent scattering, i.e.\ the Rayleigh scattering. In this way, the experiment utilizes both incoherent and coherent parts of the scattered light, for the single photon output and control, respectively. With a small modulation, the transmission signal has a large slope with zero crossing at zero detuning and is therefore ideal for the generation of an error signal. $\Delta T/T$, the error signal, is recorded with a lock-in amplifier to reject noise and the lock-in output is fed into a classical feedback scheme. The feedback output is, like the modulation, applied to the gate electrode of the device. The set-point of the control loop is the zero crossing with the goal of locking the peak of the quantum dot RF spectrum to the laser. The laser itself is locked to a HeNe laser reference.

\label{sec:Experiment}
The self-assembled InGaAs quantum dots, grown by molecular beam epitaxy, are integrated into a semiconductor charge-tunable heterostructure \cite{Warburton2000}. The quantum dots are located 25 nm above a heavily n-doped GaAs back contact ($n=1.7 \times 10^{18}$ cm$^{-3}$). The intermediate layer, undoped GaAs (25 nm), acts as a tunneling barrier. A 150 nm GaAs layer caps the quantum dots and an AlAs/GaAs superlattice (68 periods of AlAs/GaAs 3 nm/1 nm) completes the heterostructure. A Ti/Au (5 nm/10 nm) Schottky gate is deposited on the sample surface; Ohmic contacts are prepared to the back contact. Bias $V_g$ is applied between the Schottky gate and the back contact. The sample is placed in a liquid helium bath cryostat at 4.2 K with a residual magnetic field of 10 mT.

The single quantum dot spectroscopy is performed with a confocal microscope. The continuous wave laser has a short-term linewidth of 1 MHz. Long-term wavelength stability of $\sim 2$ MHz is achieved by locking the laser to a high resolution wavemeter, itself locked to a high quality HeNe laser. The size of the focal spot and the collection efficiency of the single quantum dot RF are both enhanced with a half-sphere zirconia solid immersion lens positioned on top of the Schottky gate. Fig.\ \ref{fig:1}(b) shows a RF signal from the neutral exciton transition, $\ket{0}\leftrightarrow \ket{X^{0}}$, where $\ket{X^{0}}$ represents an electron-hole complex and $\ket{0}$ the crystal ground state. The RF is detected with a silicon avalanche photodiode (APD) in single photon counting mode and the detuning of the quantum dot resonances relative to the constant frequency laser is achieved in this case with the Stark shift induced by the bias $V_{g}$. The X$^{0}$ exhibits a fine structure splitting of 11.8 $\mu$eV, the two lines having linewidths $\Gamma=1.45$, 1.28 $\mu$eV close to the transform limit of $\Gamma_0=\hbar/\tau_{\rm r} = 0.93$ $\mu$eV (220 MHz) where $\tau_r$ is the radiative lifetime of the exciton transition ($\tau_r=(0.71 \pm 0.01)$ ns here).

A sub-linewidth square-wave modulation at 527 Hz is applied to the Schottky gate. This broadens both X$^{0}$ transitions slightly, here the  ``red" transition from $\Gamma=1.45$ to $\Gamma=2.58$ $\mu$eV. The transmitted light is detected with an in situ photodiode connected to a room temperature current-voltage preamplifier. Lock-in detection of the  $\Delta T/T$ signal is shown in Fig.\ \ref{fig:1}(c). With the sub-linewidth modulation, the $\Delta T/T$ resonance is proportional to the derivative of the RF spectrum\cite{Alen2003}. There are two points which cross with high slope through zero, one for each X$^{0}$ transition. Both crossing points enable a feedback scheme: $\Delta T/T$ provides the error signal, $V_g$ the control parameter. For instance, if the transition energy increases due to electric fluctuations in the sample, $\Delta T/T$ moves away from zero. Once this is detected, a modified $V_g$ is applied to the gate to bring the resonance back to the set point. For the feedback circuit we use a PID loop. The proportional factor $P=0.1$ is chosen with respect to the slope of the error signal, while the integral $I=0.06$ and the derivative constant $D=6\times 10^{-5}$ were obtained by tuning methods. The signal:noise ratio in the $\Delta T/T$ circuit allows us to run the feedback scheme with a bandwidth up to $\sim 50$ Hz matching the frequency range of the device's charge fluctuations. The fluctuations of the nuclear spins exceed the bandwidth of the feedback\cite{andi}. The ``red" X$^{0}$ transition was used for the subsequent feedback experiments because it has a higher $\Delta T/T$ contrast than the ``blue" X$^{0}$ transition.

\begin{figure*}
\includegraphics[width=\linewidth]{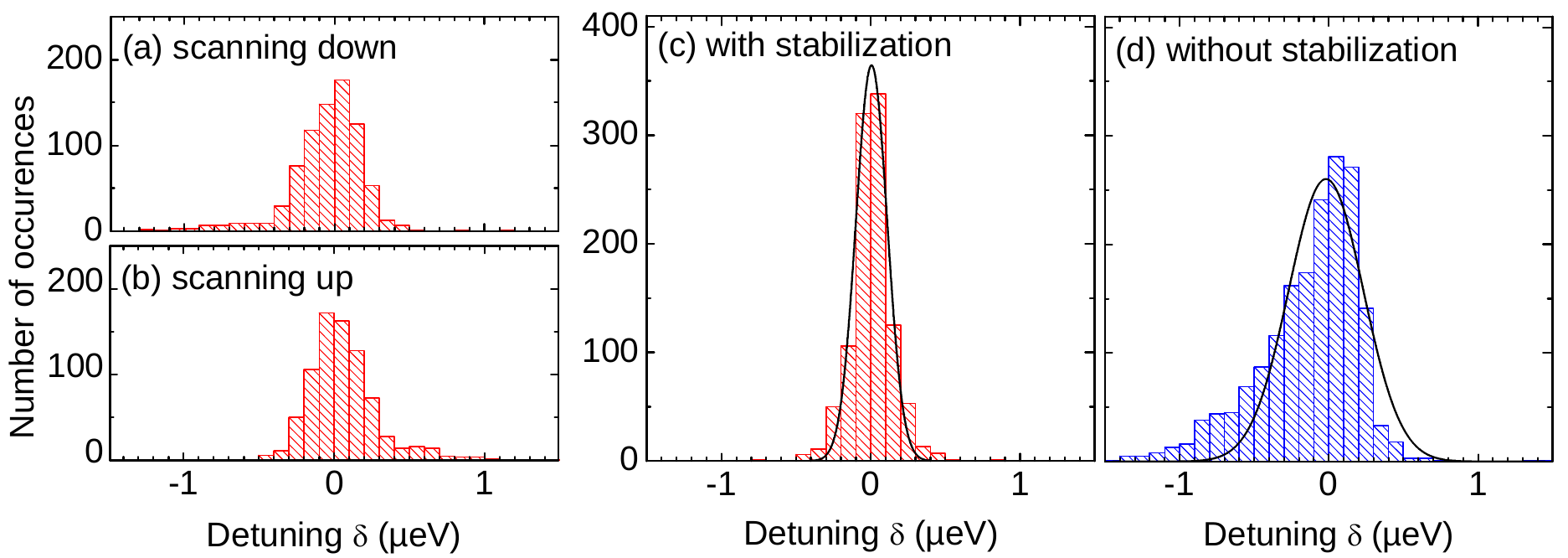}
\vspace*{-0.6cm}
\caption{
Histogram of the RF peak position with (a, b, c) and without (d) the stabilization scheme. A triangle $V_{g}$ is applied. The scanning rate of the laser is 8.0 $\mu$eV/s with period 10 s. Histograms of the RF peak position for up-sweeps (a) and down-sweeps (b) recorded with feedback.  (c) shows the histogram with feedback, negative detunings from the up-sweeps, positive detunings from the down-sweeps. A histogram without feedback is shown in (d). The standard deviation $\sigma$ is reduced from (d) 0.250 $\mu$eV (61 MHz) without active stabilization to (c) 0.089 $\mu$eV (22 MHz) with active stabilization.
}
\label{fig:2}
\end{figure*}

The performance of the single quantum dot frequency stabilization is put to the test in a stroboscopic experiment. The X$^0$ transition energy is mapped with a second laser (linewidth also \mbox{1 MHz}). The first laser stabilizes the transition with the feedback scheme at a power corresponding to a Rabi energy $\Omega$ of 0.74 $\mu$eV. A second laser of identical power is tuned with triangular function back and forth through the same transition with a rate of 8.0 $\mu$eV/s. The sum of the power of both lasers was selected to lie below the power at which power broadening becomes significant. A Lorentzian function is used to fit the data and the center position of the resonance is extracted. In this way, a ``snap-shot" of the resonance position is recorded every 5 s with ``exposure time" 100 ms for a total of 1,000 s. The distribution of the peak position can be seen in the histogram in Fig.\ \ref{fig:2}. In Fig.\ \ref{fig:2}(a) and (b), the scanning laser results in an asymmetry: the resonance frequency is more likely to lie at positive detunings on sweeping from negative to positive detunings, and vice versa. This is probably related to the so-called ``dragging" \cite{Latta2009} which is very pronounced on this quantum dot at high magnetic fields (above 0.1 T) \cite{andiRSI}: the nuclear spins polarize in such a way as to maintain the resonance with the laser over large detunings. In other words, it is likely that the asymmetries in Fig.\ \ref{fig:2}(a) and (b) are first hints of dragging. The histogram in (c) is a combination of the data sets of (a) and (b) which are influenced least by dragging (up-sweeps at negative detuning, down-sweeps at positive detuning). Without the stabilizing loop, the long term drift results in a broader distribution, Fig.\ \ref{fig:2}(d), the strong asymmetry reflecting the $1/f$-like noise, a drift to the red in this particular case. The fluctuations in resonance positions are quantified with the standard deviation $\sigma_{E}$ of the peak positions. Without stabilization Fig.\ \ref{fig:2}(d), $\sigma_{E}^{\rm OFF}=0.250$ $\mu$eV (61 MHz). With active stabilization, $\sigma=0.102$ $\mu$eV (25 MHz). This value is small enough to be influenced by shot noise in each data point which results in an energy uncertainty on fitting each spectrum to a Lorentzian. The shot noise results in an energy jitter of 0.049 $\mu$eV, giving $\sigma_{E}^{\rm ON}=0.089$ $\mu$eV (22 MHz), 36\% of $\sigma_{E}^{\rm OFF}$. The measurement of $\sigma_{E}$ represents a measurement of the noise in a bandwidth from $\sim 1$ mHz to $\sim 3.1$ Hz. (Noise at higher frequencies is reflected in the linewidth $\Gamma$.) The ratio $\sigma_{E}^{\rm OFF}:\sigma_{E}^{\rm ON}$ would increase if lower frequencies were included on account of the $1/f$-like noise.

The ultimate operation capability of the stabilization system is limited by the random noise in the output of the PID electronics. In Fig.\ \ref{fig:1}(c) the noise in the $\Delta T/T$ signal is $\sigma_{\Delta T/T}=1.45 \times 10^{-4}$. In the ideal case, this determines the energy jitter of the quantum dot resonance position\cite{nag},
\begin{equation}
\sigma_{E, {\rm min}}= \frac{d\delta}{d \Delta T} \sigma_{\Delta T/T} \simeq 0.013\; \mu{\rm eV} \; (3 \; {\rm MHz})
\end{equation}
where $\delta$ is the detuning. This limit, $\sim 100$ times smaller than the linewidth, shows the power of this technique. We have not yet reached this limit in practice. Nevertheless, stabilization with a residual jitter down to just $\sigma_f=20$ MHz is achieved.

\begin{figure}
\includegraphics[width=0.5\linewidth]{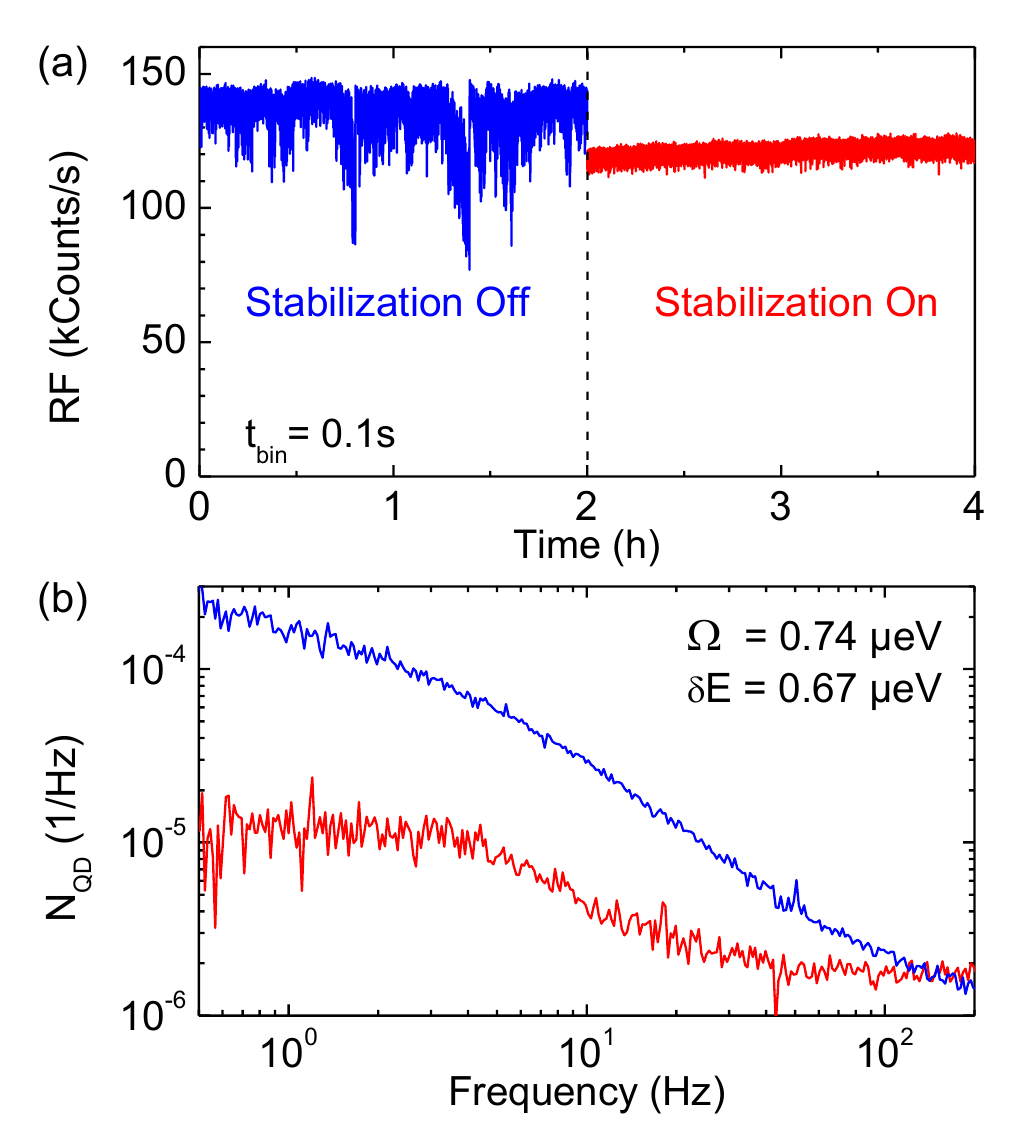}
\vspace*{-0.5cm}
\caption{(Color)
(a) Time trace of the resonance fluorescence (RF) of a single quantum dot (the one from Fig.\ \ref{fig:1}) with $\delta$ = 0 $\mu$eV recorded over several hours. The binning time was $t_{\rm bin}=100$ ms. The time trace is plotted with (red) and without (blue) the dynamic stabilization scheme. (b) Noise spectra of the normalized RF signal, $S(t)/\langle S(t)\rangle$, corresponding to the time traces of (a) after correction for external noise sources.
}
\label{fig:3}
\end{figure}

The frequency locking feedback scheme was also tested regarding its long term behaviour and bandwidth. The RF signal was recorded over several hours, Fig.\ \ref{fig:3}(a), without (blue) and with (red) the stabilizing loop. The measurements are accomplished by tuning the X$^{0}$ of the quantum dot via the Stark effect into resonance with the excitation laser ($\delta$ = 0 $\mu$eV) and then recording the arrival time of each single photon detected by the APD over the duration of the entire experiment $T$. Post-experiment, the data are analyzed by setting a binning time, $t_{\rm bin}=100$ ms in this case. For a fixed $V_{g}$, the RF counts show large fluctuations up to a factor of 2 (blue curve). The origin are slow electrical fluctuations in the sample which cause the transition to drift out of resonance with the laser. With the feedback on, these fluctuations disappear and the RF remains at a constant level (red curve). The fluctuations in the red curve arise almost entirely from shot noise in the detector. The average RF signal falls slightly with feedback because the applied modulation broadens slightly the resonance.

Insight into the bandwidth of the stabilization mechanism is revealed by a fast Fourier transform (FFT) of the time trace. Although the shot noise dominates, the shot noise can be independently measured with a small amount of reflected laser light as a source, allowing us to determine the noise coming solely from the quantum dot. The FFT of the normalized RF signal $S(t)/\langle S(t)\rangle$ provides a noise spectrum\cite{andi}:
\begin{equation}
N_{\rm RF}(f) = \vert {\rm FFT}[S(t)/\langle S(t)\rangle]\vert^2 (t_{\rm bin})^2/T.
\end{equation}
For $N_{\rm RF}(f)$, $t_{\rm bin}=1$ $\mu$s and $T = 2$ hours. The noise spectrum of the quantum dot $N_{\rm QD}(f)$ is obtained by correcting the RF noise by the noise of the experiment $N_{\rm exp}(f)$ [$N_{\rm QD}(f) = N_{\rm RF}(f) - N_{\rm exp}(f)$]. $N_{\rm QD}(f)$ corresponding to the time traces of Fig.\ \ref{fig:3}(a) are shown in Fig.\ \ref{fig:3}(b). Without feedback, $N_{\rm QD}(f)$ has a $1/f$-like dependence on $f$ as a consequence of charge noise in the device. With feedback, $N_{\rm QD}(f)$ is reduced by up to a factor of 20 at the lowest frequencies, and is constant: the $1/f$-like noise is eliminated. The two curves meet at $f \simeq 130$ Hz once the bandwidth of the PID circuit has been exceeded. At higher frequency the noise spectrum is dominated by spin noise\cite{andi}.

The quantum dot noise $N_{\rm QD}(f)$ under feedback can be linked to the jitter in the energy detuning, $\sigma_{E}$. The energy jitter is much less than the linewidth such that the change in the RF signal ($\Delta {\rm RF}$) is related quadratically to the detuning for fluctuations around $\delta=0$. The variance of the RF noise, $\sigma_{\rm RF}^{2}$, is related to an integral of the noise curve, $\sigma_{\rm RF}^{2}=\int N_{\rm QD}(f)df$\cite{Kogan}. Integrating up to frequency $\Delta f$ in the regime where $N_{\rm QD}(f)$ is approximately constant, 
\begin{equation}
\sigma_{E}^{\rm ON}=\frac{\Gamma}{2} \left(\frac{N_{\rm QD}(0) \Delta f}{3}\right)^{\frac{1}{4}}.
\end{equation}
With $\Delta f=3.1$ Hz, $N_{\rm QD}(0)=1.0 \times 10^{-5}$, $\Gamma=2.58$ $\mu$eV this predicts $\sigma_{E}^{\rm ON}=0.073$ $\mu$eV, in excellent agreement with the measurement from the stroboscopic experiment (0.089 $\mu$eV).

\begin{figure}[b]
\includegraphics[width=0.5\linewidth]{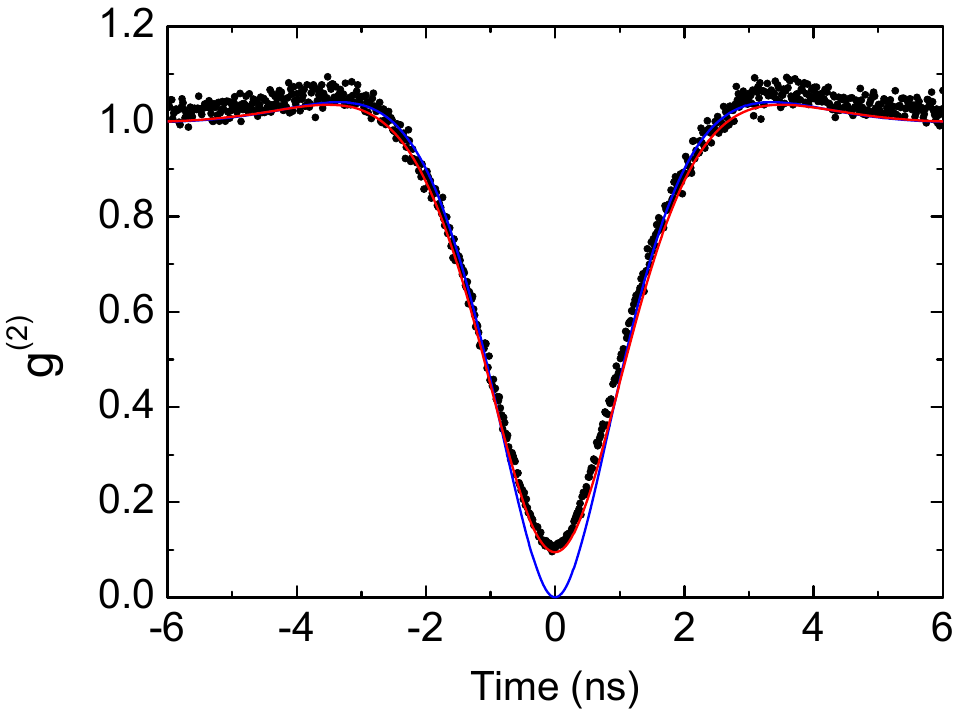}
\vspace*{-0.3cm}
\caption{(Color)
Second-order correlation $g^{2}(t)$ for the stabilized RF from the X$^{0}$ (black points). The red curve shows a convolution of the two-level atom result with a Gaussian distribution which describes the timing jitter of the detectors. The blue curve shows the two-level atom response alone.
}
\label{fig:4}
\end{figure}
An intensity correlation measurement $g^{(2)}(t)$ was performed with a Hanbury Brown-Twiss interferometer. Low noise $g^{(2)}(t)$ can only be determined at these count rates (50 kHz per APD) by integrating over several hours and the feedback is therefore important to ensure that the detuning of the quantum dot with respect to the laser remains constant. $g^{(2)}(t)$ is shown in Fig.\ \ref{fig:4} from X$^{0}$ of the same quantum dot with zero detuning. $g^{(2)}(t)$ falls to 10\% at $t=0$. This does not reflect $g^{(2)}(0)$ of the quantum dot but rather the timing jitter of the detectors which is comparable to the radiative lifetime. We attempt to describe $g^{(2)}(t)$ with a convolution of $g^{(2)}(t)$ for an ideal two-level atom, $g^{(2)}_{\rm atom}(t)$, and the response of the detectors $G(t)$:
\begin{equation}
g^{(2)}(t) = g^{(2)}_{\rm atom}(t) \otimes G(t).
\end{equation}
The detector response is a Gaussian function,
\begin{equation}
G(t)=\frac{1}{\sqrt{2 \pi}\sigma_{D}} \exp\left(-\frac{t^2}{2\sigma_{D}^{2}}\right).
\end{equation}
$g^{(2)}_{\rm atom}(t)$ of a 2-level system with resonant excitation is, 
\begin{equation}
g^{(2)}_{\rm atom}(t)= 
1-\left[\cos(\lambda t)+ \frac{3}{4\tau_r} \lambda \sin (\lambda t)\right] \exp\left(-\frac{3t}{4\tau_r}\right)
\end{equation}
with $\lambda = (\Omega^2-(1/4\tau_r)^2)^{1/2}$. 
The temporal jitter of the detector $\tau_D=0.40$ ns is measured independently. $\Omega$ and $\tau_r$ are known from other experiments to within $10-20$\% and are allowed to vary in these windows by a fit routine. The convolution provides an excellent description of the measured $g^{(2)}(t)$ with $\Omega= (0.99\pm 0.1)$ $\mu$eV and $\tau_r= (0.78\pm 0.05)$ ns. In particular, with low systematic error we can set an upper bound to the quantum dot $g^{(2)}(0)$ of 1-2\%.

\label{sec:Conclusion}
In conclusion, we have developed a dynamic method of locking the optical resonance of a single quantum dot to a stabilized laser in order to produce a stream of frequency-stabilized single photons via resonance fluorescence. Generally speaking, the scheme represents a way to reduce the local charge noise in a semiconductor. 

Now that the basic principle is established, there are options for improving the feedback scheme. First, the remaining jitter in the quantum dot resonance position can be reduced by reducing the noise in the transmission detection. Presently, we are far from the limit defined by the shot noise in the detector current. With lower noise, the feedback bandwidth can also be increased. The tantalizing prospect is to create transform-limited linewidths routinely with high bandwidth feedback. A bandwidth of about 50 kHz is required \cite{andi}. Secondly, the modulation required here to generate the error signal could be eliminated in a number of ways. For instance, a dispersive lineshape can arise naturally in reflectivity via weak coupling to a cavity \cite{Alen2006}; or the Faraday effect in a small magnetic field \cite{Atatuere2007} could be used.

We acknowledge support from the Swiss National Science Foundation (SNF) and NCCR QSIT.
A.L., D.R. and A.D.W. acknowledge gratefully support from DFG SPP1285 and BMBF QuaHLRep 01BQ1035.


\begin{thebibliography}{33}%
\makeatletter
\providecommand \@ifxundefined [1]{%
 \@ifx{#1\undefined}
}%
\providecommand \@ifnum [1]{%
 \ifnum #1\expandafter \@firstoftwo
 \else \expandafter \@secondoftwo
 \fi
}%
\providecommand \@ifx [1]{%
 \ifx #1\expandafter \@firstoftwo
 \else \expandafter \@secondoftwo
 \fi
}%
\providecommand \natexlab [1]{#1}%
\providecommand \enquote  [1]{``#1''}%
\providecommand \bibnamefont  [1]{#1}%
\providecommand \bibfnamefont [1]{#1}%
\providecommand \citenamefont [1]{#1}%
\providecommand \href@noop [0]{\@secondoftwo}%
\providecommand \href [0]{\begingroup \@sanitize@url \@href}%
\providecommand \@href[1]{\@@startlink{#1}\@@href}%
\providecommand \@@href[1]{\endgroup#1\@@endlink}%
\providecommand \@sanitize@url [0]{\catcode `\\12\catcode `\$12\catcode
  `\&12\catcode `\#12\catcode `\^12\catcode `\_12\catcode `\%12\relax}%
\providecommand \@@startlink[1]{}%
\providecommand \@@endlink[0]{}%
\providecommand \url  [0]{\begingroup\@sanitize@url \@url }%
\providecommand \@url [1]{\endgroup\@href {#1}{\urlprefix }}%
\providecommand \urlprefix  [0]{URL }%
\providecommand \Eprint [0]{\href }%
\providecommand \doibase [0]{http://dx.doi.org/}%
\providecommand \selectlanguage [0]{\@gobble}%
\providecommand \bibinfo  [0]{\@secondoftwo}%
\providecommand \bibfield  [0]{\@secondoftwo}%
\providecommand \translation [1]{[#1]}%
\providecommand \BibitemOpen [0]{}%
\providecommand \bibitemStop [0]{}%
\providecommand \bibitemNoStop [0]{.\EOS\space}%
\providecommand \EOS [0]{\spacefactor3000\relax}%
\providecommand \BibitemShut  [1]{\csname bibitem#1\endcsname}%
\let\auto@bib@innerbib\@empty
\bibitem [{\citenamefont {Lounis}\ and\ \citenamefont
  {Orrit}(2005)}]{Lounis2005}%
  \BibitemOpen
  \bibfield  {author} {\bibinfo {author} {\bibfnamefont {B.}~\bibnamefont
  {Lounis}}\ and\ \bibinfo {author} {\bibfnamefont {M.}~\bibnamefont {Orrit}},\
  }\href {\doibase 10.1088/0034-4885/68/5/R04}{\emph {\bibinfo {title} {Single-Photon  Sources}}},  {\bibfield  {journal} {\bibinfo
  {journal} {Rep. Prog. Phys.}\ }\textbf {\bibinfo {volume}
  {68}},\ \bibinfo {pages} {1129} (\bibinfo {year} {2005})}\BibitemShut
  {NoStop}%
\bibitem [{\citenamefont {Michler}\ \emph {et~al.}(2000)\citenamefont
  {Michler}, \citenamefont {Kiraz}, \citenamefont {Becher}, \citenamefont
  {Schoenfeld}, \citenamefont {Petroff}, \citenamefont {Zhang}, \citenamefont
  {Hu},\ and\ \citenamefont {Imamo\u{g}lu}}]{Michler2000}%
  \BibitemOpen
  \bibfield  {author} {\bibinfo {author} {\bibfnamefont {P.}~\bibnamefont
  {Michler}}, \bibinfo {author} {\bibfnamefont {A.}~\bibnamefont {Kiraz}},
  \bibinfo {author} {\bibfnamefont {C.}~\bibnamefont {Becher}}, \bibinfo
  {author} {\bibfnamefont {W.~V.}\ \bibnamefont {Schoenfeld}}, \bibinfo
  {author} {\bibfnamefont {P.~M.}\ \bibnamefont {Petroff}}, \bibinfo {author}
  {\bibfnamefont {L.~D.}\ \bibnamefont {Zhang}}, \bibinfo {author}
  {\bibfnamefont {E.}~\bibnamefont {Hu}}, \ and\ \bibinfo {author}
  {\bibfnamefont {A.}~\bibnamefont {Imamo\u{g}lu}},\ }\href {\doibase
  10.1126/science.290.5500.2282} {\emph {\bibinfo {title} {A Quantum Dot Single-Photon Turnstile Device}}}, {\bibfield  {journal} {\bibinfo  {journal}
  {Science}\ }\textbf {\bibinfo {volume} {290}},\ \bibinfo {pages} {2282}
  (\bibinfo {year} {2000})}\BibitemShut {NoStop}%
\bibitem [{\citenamefont {Santori}\ \emph {et~al.}(2002)\citenamefont
  {Santori}, \citenamefont {Fattal}, \citenamefont {Vu\u{c}kovi\'c}, \citenamefont
  {Solomon},\ and\ \citenamefont {Yamamoto}}]{Santori2002}%
  \BibitemOpen
  \bibfield  {author} {\bibinfo {author} {\bibfnamefont {C.}~\bibnamefont
  {Santori}}, \bibinfo {author} {\bibfnamefont {D.}~\bibnamefont {Fattal}},
  \bibinfo {author} {\bibfnamefont {J.}~\bibnamefont {Vu\u{c}kovi\'c}}, \bibinfo
  {author} {\bibfnamefont {G.~S.}\ \bibnamefont {Solomon}}, \ and\ \bibinfo
  {author} {\bibfnamefont {Y.}~\bibnamefont {Yamamoto}},\ }\href {\doibase
  10.1038/nature01086} {\emph {\bibinfo {title} {Indistinguishable Photons from a Single-Photon Device}}}, {\bibfield  {journal} {\bibinfo  {journal} {Nature (London)}\
  }\textbf {\bibinfo {volume} {419}},\ \bibinfo {pages} {594} (\bibinfo {year}
  {2002})}\BibitemShut {NoStop}%
\bibitem [{\citenamefont {Shields}(2007)}]{Shields2007}%
  \BibitemOpen
  \bibfield  {author} {\bibinfo {author} {\bibfnamefont {A.~J.}\ \bibnamefont
  {Shields}},\ }\href {\doibase 10.1038/nphoton.2007.46} {\emph {\bibinfo {title} {Semiconductor Quantum Light Sources}}}, {\bibfield  {journal}
  {\bibinfo  {journal} {Nat. Phot.}\ }\textbf {\bibinfo {volume} {1}},\
  \bibinfo {pages} {215} (\bibinfo {year} {2007})}\BibitemShut {NoStop}%
\bibitem [{\citenamefont {Muller}\ \emph {et~al.}(2007)\citenamefont {Muller},
  \citenamefont {Flagg}, \citenamefont {Bianucci}, \citenamefont {Wang},
  \citenamefont {Deppe}, \citenamefont {Ma}, \citenamefont {Zhang},
  \citenamefont {Salamo}, \citenamefont {Xiao},\ and\ \citenamefont
  {Shih}}]{Muller2007}%
  \BibitemOpen
  \bibfield  {author} {\bibinfo {author} {\bibfnamefont {A.}~\bibnamefont
  {Muller}}, \bibinfo {author} {\bibfnamefont {E.~B.}\ \bibnamefont {Flagg}},
  \bibinfo {author} {\bibfnamefont {P.}~\bibnamefont {Bianucci}}, \bibinfo
  {author} {\bibfnamefont {X.~Y.}\ \bibnamefont {Wang}}, \bibinfo {author}
  {\bibfnamefont {D.~G.}\ \bibnamefont {Deppe}}, \bibinfo {author}
  {\bibfnamefont {W.}~\bibnamefont {Ma}}, \bibinfo {author} {\bibfnamefont
  {J.}~\bibnamefont {Zhang}}, \bibinfo {author} {\bibfnamefont {G.~J.}\
  \bibnamefont {Salamo}}, \bibinfo {author} {\bibfnamefont {M.}~\bibnamefont
  {Xiao}}, \ and\ \bibinfo {author} {\bibfnamefont {C.~K.}\ \bibnamefont
  {Shih}},\ }\href {\doibase 10.1103/PhysRevLett.99.187402} {\emph {\bibinfo {title} {Resonance Fluorescence from a Coherently Driven Semiconductor Quantum Dot in a Cavity}}}, {\bibfield
  {journal} {\bibinfo  {journal} {Phys. Rev. Lett.}\ }\textbf {\bibinfo
  {volume} {99}},\ \bibinfo {pages} {187402} (\bibinfo {year}
  {2007})}\BibitemShut {NoStop}%
\bibitem [{\citenamefont {Dalgarno}\ \emph {et~al.}(2008)\citenamefont
  {Dalgarno}, \citenamefont {Smith}, \citenamefont {McFarlane}, \citenamefont
  {Gerardot}, \citenamefont {Karrai}, \citenamefont {Badolato}, \citenamefont
  {Petroff},\ and\ \citenamefont {Warburton}}]{Dalgarno2008a}%
  \BibitemOpen
  \bibfield  {author} {\bibinfo {author} {\bibfnamefont {P.~A.}\ \bibnamefont
  {Dalgarno}}, \bibinfo {author} {\bibfnamefont {J.~M.}\ \bibnamefont {Smith}},
  \bibinfo {author} {\bibfnamefont {J.}~\bibnamefont {McFarlane}}, \bibinfo
  {author} {\bibfnamefont {B.~D.}\ \bibnamefont {Gerardot}}, \bibinfo {author}
  {\bibfnamefont {K.}~\bibnamefont {Karrai}}, \bibinfo {author} {\bibfnamefont
  {A.}~\bibnamefont {Badolato}}, \bibinfo {author} {\bibfnamefont {P.~M.}\
  \bibnamefont {Petroff}}, \ and\ \bibinfo {author} {\bibfnamefont {R.~J.}\
  \bibnamefont {Warburton}},\ }\href {\doibase 10.1103/PhysRevB.77.245311}
  {\emph {\bibinfo {title} {Coulomb Interactions in Single Charged Self-Assembled Quantum Dots: Radiative Lifetime and Recombination Energy}}}, {\bibfield  {journal} {\bibinfo  {journal} {Phys. Rev. B}\ }\textbf {\bibinfo
  {volume} {77}},\ \bibinfo {pages} {245311} (\bibinfo {year}
  {2008})}\BibitemShut {NoStop}%
\bibitem [{\citenamefont {Vahala}(2003)}]{Vahala2003}%
  \BibitemOpen
  \bibfield  {author} {\bibinfo {author} {\bibfnamefont {K.~J.}\ \bibnamefont
  {Vahala}},\ }\href@noop {} {\emph {\bibinfo {title} {Optical Microcavities}}}, {\bibfield  {journal} {\bibinfo  {journal}
  {Nature (London)}\ }\textbf {\bibinfo {volume} {424}},\ \bibinfo {pages} {839}
  (\bibinfo {year} {2003})}\BibitemShut {NoStop}%
\bibitem [{\citenamefont {Claudon}\ \emph {et~al.}(2010)\citenamefont
  {Claudon}, \citenamefont {Bleuse}, \citenamefont {Malik}, \citenamefont
  {Bazin}, \citenamefont {Jaffrennou}, \citenamefont {Gregersen}, \citenamefont
  {Sauvan}, \citenamefont {Lalanne},\ and\ \citenamefont
  {Gérard}}]{Claudon2010}%
  \BibitemOpen
  \bibfield  {author} {\bibinfo {author} {\bibfnamefont {J.}~\bibnamefont
  {Claudon}}, \bibinfo {author} {\bibfnamefont {J.}~\bibnamefont {Bleuse}},
  \bibinfo {author} {\bibfnamefont {N.~S.}\ \bibnamefont {Malik}}, \bibinfo
  {author} {\bibfnamefont {M.}~\bibnamefont {Bazin}}, \bibinfo {author}
  {\bibfnamefont {P.}~\bibnamefont {Jaffrennou}}, \bibinfo {author}
  {\bibfnamefont {N.}~\bibnamefont {Gregersen}}, \bibinfo {author}
  {\bibfnamefont {C.}~\bibnamefont {Sauvan}}, \bibinfo {author} {\bibfnamefont
  {P.}~\bibnamefont {Lalanne}}, \ and\ \bibinfo {author} {\bibfnamefont
  {J.-M.}\ \bibnamefont {G\'erard}},\ }\href {\doibase 10.1038/nphoton.2009.287}
  {\emph {\bibinfo {title} {A Highly Efficient Single-Photon Source Based on a Quantum Dot in a Photonic Nanowire}}}, {\bibfield  {journal} {\bibinfo  {journal} {Nat. Phot.}\ }\textbf
  {\bibinfo {volume} {4}},\ \bibinfo {pages} {174} (\bibinfo {year}
  {2010})}\BibitemShut {NoStop}%
\bibitem [{\citenamefont {H\"ogele}\ \emph {et~al.}(2004)\citenamefont {Hogele},
  \citenamefont {Seidl}, \citenamefont {Kroner}, \citenamefont {Karrai},
  \citenamefont {Warburton}, \citenamefont {Gerardot},\ and\ \citenamefont
  {Petroff}}]{Hogele2004}%
  \BibitemOpen
  \bibfield  {author} {\bibinfo {author} {\bibfnamefont {A.}~\bibnamefont
  {H\"ogele}}, \bibinfo {author} {\bibfnamefont {S.}~\bibnamefont {Seidl}},
  \bibinfo {author} {\bibfnamefont {M.}~\bibnamefont {Kroner}}, \bibinfo
  {author} {\bibfnamefont {K.}~\bibnamefont {Karrai}}, \bibinfo {author}
  {\bibfnamefont {R.~J.}\ \bibnamefont {Warburton}}, \bibinfo {author}
  {\bibfnamefont {B.~D.}\ \bibnamefont {Gerardot}}, \ and\ \bibinfo {author}
  {\bibfnamefont {P.~M.}\ \bibnamefont {Petroff}},\ }\href {\doibase
  10.1103/PhysRevLett.93.217401} {\emph {\bibinfo {title} {Voltage-Controlled Optics of a Quantum Dot}}}, {\bibfield  {journal} {\bibinfo  {journal}
  {Phys. Rev. Lett.}\ }\textbf {\bibinfo {volume} {93}},\ \bibinfo {pages}
  {217401} (\bibinfo {year} {2004})}\BibitemShut {NoStop}%
\bibitem [{\citenamefont {Atat\"ure}\ \emph {et~al.}(2006)\citenamefont
  {Atat\"ure}, \citenamefont {Dreiser}, \citenamefont {Badolato}, \citenamefont
  {H\"ogele}, \citenamefont {Karrai},\ and\ \citenamefont
  {Imamoglu}}]{Atature2006}%
  \BibitemOpen
  \bibfield  {author} {\bibinfo {author} {\bibfnamefont {M.}~\bibnamefont
  {Atat\"ure}}, \bibinfo {author} {\bibfnamefont {J.}~\bibnamefont {Dreiser}},
  \bibinfo {author} {\bibfnamefont {A.}~\bibnamefont {Badolato}}, \bibinfo
  {author} {\bibfnamefont {A.}~\bibnamefont {H\"ogele}}, \bibinfo {author}
  {\bibfnamefont {K.}~\bibnamefont {Karrai}}, \ and\ \bibinfo {author}
  {\bibfnamefont {A.}~\bibnamefont {Imamoglu}},\ }\href {\doibase
  10.1126/science.1126074} {\emph {\bibinfo {title} {Quantum-Dot Spin-State Preparation with Near-Unity Fidelity}}}, {\bibfield  {journal} {\bibinfo  {journal}
  {Science}\ }\textbf {\bibinfo {volume} {312}},\ \bibinfo {pages} {551}
  (\bibinfo {year} {2006})}\BibitemShut {NoStop}%
\bibitem [{\citenamefont {Houel}\ \emph {et~al.}(2012)\citenamefont {Houel},
  \citenamefont {Kuhlmann}, \citenamefont {Greuter}, \citenamefont {Xue},
  \citenamefont {Poggio}, \citenamefont {Gerardot}, \citenamefont {Dalgarno},
  \citenamefont {Badolato}, \citenamefont {Petroff}, \citenamefont {Ludwig},
  \citenamefont {Reuter}, \citenamefont {Wieck},\ and\ \citenamefont
  {Warburton}}]{Houel2012}%
  \BibitemOpen
  \bibfield  {author} {\bibinfo {author} {\bibfnamefont {J.}~\bibnamefont
  {Houel}}, \bibinfo {author} {\bibfnamefont {A.~V.}\ \bibnamefont {Kuhlmann}},
  \bibinfo {author} {\bibfnamefont {L.}~\bibnamefont {Greuter}}, \bibinfo
  {author} {\bibfnamefont {F.}~\bibnamefont {Xue}}, \bibinfo {author}
  {\bibfnamefont {M.}~\bibnamefont {Poggio}}, \bibinfo {author} {\bibfnamefont
  {B.~D.}\ \bibnamefont {Gerardot}}, \bibinfo {author} {\bibfnamefont {P.~A.}\
  \bibnamefont {Dalgarno}}, \bibinfo {author} {\bibfnamefont {A.}~\bibnamefont
  {Badolato}}, \bibinfo {author} {\bibfnamefont {P.~M.}\ \bibnamefont
  {Petroff}}, \bibinfo {author} {\bibfnamefont {A.}~\bibnamefont {Ludwig}},
  \bibinfo {author} {\bibfnamefont {D.}~\bibnamefont {Reuter}}, \bibinfo
  {author} {\bibfnamefont {A.~D.}\ \bibnamefont {Wieck}}, \ and\ \bibinfo
  {author} {\bibfnamefont {R.~J.}\ \bibnamefont {Warburton}},\ }\href {\doibase
  10.1103/PhysRevLett.108.107401} {\emph {\bibinfo {title} {Probing Single-Charge Fluctuations at a $\mathrm{GaAs}/\mathrm{AlAs}$ Interface Using Laser Spectroscopy on a Nearby InGaAs Quantum Dot}}}, {\bibfield  {journal} {\bibinfo  {journal}
  {Phys. Rev. Lett.}\ }\textbf {\bibinfo {volume} {108}},\ \bibinfo {pages}
  {107401} (\bibinfo {year} {2012})}\BibitemShut {NoStop}%
\bibitem [{\citenamefont {Kuhlmann}\ \emph
  {et~al.}(2013{\natexlab{a}})\citenamefont {Kuhlmann}, \citenamefont {Houel},
  \citenamefont {Ludwig}, \citenamefont {Greuter}, \citenamefont {D.Reuter},
  \citenamefont {Wieck}, \citenamefont {Poggio},\ and\ \citenamefont
  {Warburton}}]{andi}%
  \BibitemOpen
  \bibfield  {author} {\bibinfo {author} {\bibfnamefont {A.~V.}\ \bibnamefont
  {Kuhlmann}}, \bibinfo {author} {\bibfnamefont {J.}~\bibnamefont {Houel}},
  \bibinfo {author} {\bibfnamefont {A.}~\bibnamefont {Ludwig}}, \bibinfo
  {author} {\bibfnamefont {L.}~\bibnamefont {Greuter}}, \bibinfo {author}
  {\bibnamefont {D.Reuter}}, \bibinfo {author} {\bibfnamefont {A.~D.}\
  \bibnamefont {Wieck}}, \bibinfo {author} {\bibfnamefont {M.}~\bibnamefont
  {Poggio}}, \ and\ \bibinfo {author} {\bibfnamefont {R.~J.}\ \bibnamefont
  {Warburton}},\ }\href@noop {} {\emph {\bibinfo {title} {Charge Noise and Spin Noise in a Semiconductor Quantum Device}}}, {\bibfield  {journal} {\bibinfo  {journal}
  {ArXiv:1301.6381}\ } (\bibinfo {year} {2013}{\natexlab{a}})}\BibitemShut
  {NoStop}%
\bibitem [{\citenamefont {Warburton}(2013)}]{Warburton2013}%
  \BibitemOpen
  \bibfield  {author} {\bibinfo {author} {\bibfnamefont {R.~J.}\ \bibnamefont
  {Warburton}},\ }\href {\doibase http://dx.doi.org/10.1038/nmat3585} {\emph {\bibinfo {title} {Single Spins in Self-Assembled Quantum Dots}}}, 
  {\bibfield  {journal} {\bibinfo  {journal} {Nat. Mater.}\ }\textbf
  {\bibinfo {volume} {12}},\ \bibinfo {pages} {483 } (\bibinfo {year}
  {2013})}\BibitemShut {NoStop}%
\bibitem [{\citenamefont {Yilmaz}\ \emph {et~al.}(2010)\citenamefont {Yilmaz},
  \citenamefont {Fallahi},\ and\ \citenamefont {Imamo\u{g}lu}}]{Yilmaz2010}%
  \BibitemOpen
  \bibfield  {author} {\bibinfo {author} {\bibfnamefont {S.~T.}\ \bibnamefont
  {Yilmaz}}, \bibinfo {author} {\bibfnamefont {P.}~\bibnamefont {Fallahi}}, \
  and\ \bibinfo {author} {\bibfnamefont {A.}~\bibnamefont {Imamo\u{g}lu}},\ }\href
  {\doibase 10.1103/PhysRevLett.105.033601} {\emph {\bibinfo {title} {Quantum-Dot-Spin Single-Photon Interface}}}, {\bibfield  {journal} {\bibinfo
  {journal} {Phys. Rev. Lett.}\ }\textbf {\bibinfo {volume} {105}},\ \bibinfo
  {pages} {033601} (\bibinfo {year} {2010})}\BibitemShut {NoStop}%
\bibitem [{\citenamefont {De~Greve}\ \emph {et~al.}(2012)\citenamefont
  {De~Greve}, \citenamefont {Yu}, \citenamefont {McMahon}, \citenamefont
  {Pelc}, \citenamefont {Natarajan}, \citenamefont {Kim}, \citenamefont {Abe},
  \citenamefont {Maier}, \citenamefont {Schneider}, \citenamefont {Kamp},
  \citenamefont {H\"ofling}, \citenamefont {Hadfield}, \citenamefont {Forchel},
  \citenamefont {Fejer},\ and\ \citenamefont {Yamamoto}}]{DeGreve2012}%
  \BibitemOpen
  \bibfield  {author} {\bibinfo {author} {\bibfnamefont {K.}~\bibnamefont
  {De~Greve}}, \bibinfo {author} {\bibfnamefont {L.}~\bibnamefont {Yu}},
  \bibinfo {author} {\bibfnamefont {P.~L.}\ \bibnamefont {McMahon}}, \bibinfo
  {author} {\bibfnamefont {J.~S.}\ \bibnamefont {Pelc}}, \bibinfo {author}
  {\bibfnamefont {C.~M.}\ \bibnamefont {Natarajan}}, \bibinfo {author}
  {\bibfnamefont {N.~Y.}\ \bibnamefont {Kim}}, \bibinfo {author} {\bibfnamefont
  {E.}~\bibnamefont {Abe}}, \bibinfo {author} {\bibfnamefont {S.}~\bibnamefont
  {Maier}}, \bibinfo {author} {\bibfnamefont {C.}~\bibnamefont {Schneider}},
  \bibinfo {author} {\bibfnamefont {M.}~\bibnamefont {Kamp}}, \bibinfo {author}
  {\bibfnamefont {S.}~\bibnamefont {H\"ofling}}, \bibinfo {author}
  {\bibfnamefont {R.~H.}\ \bibnamefont {Hadfield}}, \bibinfo {author}
  {\bibfnamefont {A.}~\bibnamefont {Forchel}}, \bibinfo {author} {\bibfnamefont
  {M.~M.}\ \bibnamefont {Fejer}}, \ and\ \bibinfo {author} {\bibfnamefont
  {Y.}~\bibnamefont {Yamamoto}},\ }\href {\doibase 10.1038/nature11577}
  {\emph {\bibinfo {title} {Quantum-Dot Spin-Photon Entanglement via Frequency Downconversion to Telecom Wavelength}}}, {\bibfield  {journal} {\bibinfo  {journal} {Nature (London)}\ }\textbf {\bibinfo
  {volume} {491}},\ \bibinfo {pages} {421} (\bibinfo {year}
  {2012})}\BibitemShut {NoStop}%
\bibitem [{\citenamefont {Gao}\ \emph {et~al.}(2012)\citenamefont {Gao},
  \citenamefont {Fallahi}, \citenamefont {Togan}, \citenamefont
  {Miguel-Sanchez},\ and\ \citenamefont {Imamoglu}}]{Gao2012}%
  \BibitemOpen
  \bibfield  {author} {\bibinfo {author} {\bibfnamefont {W.~B.}\ \bibnamefont
  {Gao}}, \bibinfo {author} {\bibfnamefont {P.}~\bibnamefont {Fallahi}},
  \bibinfo {author} {\bibfnamefont {E.}~\bibnamefont {Togan}}, \bibinfo
  {author} {\bibfnamefont {J.}~\bibnamefont {Miguel-Sanchez}}, \ and\ \bibinfo
  {author} {\bibfnamefont {A.}~\bibnamefont {Imamoglu}},\ }\href {\doibase
  10.1038/nature11573} {\emph {\bibinfo {title} {Observation of Entanglement Between a Quantum Dot Spin and a Single Photon}}}, {\bibfield  {journal} {\bibinfo  {journal} {Nature (London)}\
  }\textbf {\bibinfo {volume} {491}},\ \bibinfo {pages} {426} (\bibinfo {year}
  {2012})}\BibitemShut {NoStop}%
\bibitem [{\citenamefont {Warburton}\ \emph {et~al.}(2002)\citenamefont
  {Warburton}, \citenamefont {Schulhauser}, \citenamefont {Haft}, \citenamefont
  {Sch\"aflein}, \citenamefont {Karrai}, \citenamefont {Garcia}, \citenamefont
  {Schoenfeld},\ and\ \citenamefont {Petroff}}]{Warburton2002}%
  \BibitemOpen
  \bibfield  {author} {\bibinfo {author} {\bibfnamefont {R.~J.}\ \bibnamefont
  {Warburton}}, \bibinfo {author} {\bibfnamefont {C.}~\bibnamefont
  {Schulhauser}}, \bibinfo {author} {\bibfnamefont {D.}~\bibnamefont {Haft}},
  \bibinfo {author} {\bibfnamefont {C.}~\bibnamefont {Sch\"aflein}}, \bibinfo
  {author} {\bibfnamefont {K.}~\bibnamefont {Karrai}}, \bibinfo {author}
  {\bibfnamefont {J.~M.}\ \bibnamefont {Garcia}}, \bibinfo {author}
  {\bibfnamefont {W.}~\bibnamefont {Schoenfeld}}, \ and\ \bibinfo {author}
  {\bibfnamefont {P.~M.}\ \bibnamefont {Petroff}},\ }\href {\doibase
  10.1103/PhysRevB.65.113303} {\emph {\bibinfo {title} {Giant Permanent Dipole Moments of Excitons in Semiconductor Nanostructures}}}, {\bibfield  {journal} {\bibinfo  {journal} {Phys.
  Rev. B}\ }\textbf {\bibinfo {volume} {65}},\ \bibinfo {pages} {113303} (\bibinfo {year} {2002})} \BibitemShut {NoStop}%
\bibitem [{\citenamefont {Bennett}\ \emph {et~al.}(2010)\citenamefont
  {Bennett}, \citenamefont {Pooley}, \citenamefont {Stevenson}, \citenamefont
  {Ward}, \citenamefont {Patel}, \citenamefont {de~la Giroday}, \citenamefont
  {Skoeld}, \citenamefont {Farrer}, \citenamefont {Nicoll}, \citenamefont
  {Ritchie},\ and\ \citenamefont {Shields}}]{Bennett2010}%
  \BibitemOpen
  \bibfield  {author} {\bibinfo {author} {\bibfnamefont {A.~J.}\ \bibnamefont
  {Bennett}}, \bibinfo {author} {\bibfnamefont {M.~A.}\ \bibnamefont {Pooley}},
  \bibinfo {author} {\bibfnamefont {R.~M.}\ \bibnamefont {Stevenson}}, \bibinfo
  {author} {\bibfnamefont {M.~B.}\ \bibnamefont {Ward}}, \bibinfo {author}
  {\bibfnamefont {R.~B.}\ \bibnamefont {Patel}}, \bibinfo {author}
  {\bibfnamefont {A.~B.}\ \bibnamefont {de~la Giroday}}, \bibinfo {author}
  {\bibfnamefont {N.}~\bibnamefont {Sk\"old}}, \bibinfo {author} {\bibfnamefont
  {I.}~\bibnamefont {Farrer}}, \bibinfo {author} {\bibfnamefont {C.~A.}\
  \bibnamefont {Nicoll}}, \bibinfo {author} {\bibfnamefont {D.~A.}\
  \bibnamefont {Ritchie}}, \ and\ \bibinfo {author} {\bibfnamefont {A.~J.}\
  \bibnamefont {Shields}},\ }\href {\doibase 10.1038/NPHYS1780} {\emph {\bibinfo {title} {Electric-Field-Induced Coherent Coupling of the Exciton States in a Single Quantum Dot}}}, {\bibfield
  {journal} {\bibinfo  {journal} {Nat. Phys.}\ }\textbf {\bibinfo {volume}
  {6}},\ \bibinfo {pages} {947} (\bibinfo {year} {2010})}\BibitemShut {NoStop}%
\bibitem [{\citenamefont {Seidl}\ \emph {et~al.}(2006)\citenamefont {Seidl},
  \citenamefont {Kroner}, \citenamefont {H\"ogele}, \citenamefont {Karrai},
  \citenamefont {Warburton}, \citenamefont {Badolato},\ and\ \citenamefont
  {Petroff}}]{Seidl2006a}%
  \BibitemOpen
  \bibfield  {author} {\bibinfo {author} {\bibfnamefont {S.}~\bibnamefont
  {Seidl}}, \bibinfo {author} {\bibfnamefont {M.}~\bibnamefont {Kroner}},
  \bibinfo {author} {\bibfnamefont {A.}~\bibnamefont {H\"ogele}}, \bibinfo
  {author} {\bibfnamefont {K.}~\bibnamefont {Karrai}}, \bibinfo {author}
  {\bibfnamefont {R.~J.}\ \bibnamefont {Warburton}}, \bibinfo {author}
  {\bibfnamefont {A.}~\bibnamefont {Badolato}}, \ and\ \bibinfo {author}
  {\bibfnamefont {P.~M.}\ \bibnamefont {Petroff}},\ }\href {\doibase
  10.1063/1.2204843} {\emph {\bibinfo {title} {Effect of Uniaxial Stress on Excitons in a Self-Assembled Quantum Dot}}}, {\bibfield  {journal} {\bibinfo  {journal} {Appl. Phys.
  Lett.}\ }\textbf {\bibinfo {volume} {88}},\ \bibinfo {pages} {203113}
  (\bibinfo {year} {2006})}\BibitemShut {NoStop}%
\bibitem [{\citenamefont {J\"ons}\ \emph {et~al.}(2011)\citenamefont {Joens},
  \citenamefont {Hafenbrak}, \citenamefont {Singh}, \citenamefont {Ding},
  \citenamefont {Plumhof}, \citenamefont {Rastelli}, \citenamefont {Schmidt},
  \citenamefont {Bester},\ and\ \citenamefont {Michler}}]{Joens2011}%
  \BibitemOpen
  \bibfield  {author} {\bibinfo {author} {\bibfnamefont {K.~D.}\ \bibnamefont
  {J\"ons}}, \bibinfo {author} {\bibfnamefont {R.}~\bibnamefont {Hafenbrak}},
  \bibinfo {author} {\bibfnamefont {R.}~\bibnamefont {Singh}}, \bibinfo
  {author} {\bibfnamefont {F.}~\bibnamefont {Ding}}, \bibinfo {author}
  {\bibfnamefont {J.~D.}\ \bibnamefont {Plumhof}}, \bibinfo {author}
  {\bibfnamefont {A.}~\bibnamefont {Rastelli}}, \bibinfo {author}
  {\bibfnamefont {O.~G.}\ \bibnamefont {Schmidt}}, \bibinfo {author}
  {\bibfnamefont {G.}~\bibnamefont {Bester}}, \ and\ \bibinfo {author}
  {\bibfnamefont {P.}~\bibnamefont {Michler}},\ }\href {\doibase
  10.1103/PhysRevLett.107.217402} {\emph {\bibinfo {title} {Dependence of the Redshifted and Blueshifted Photoluminescence Spectra of Single ${\mathrm{In}}_{x}{\mathrm{Ga}}_{1-x}\mathrm{As}/\mathrm{GaAs}$ Quantum Dots on the Applied Uniaxial Stress}}}, {\bibfield  {journal} {\bibinfo  {journal}
  {Phys. Rev. Lett.}\ }\textbf {\bibinfo {volume} {107}},\ \bibinfo {pages}
  {217402} (\bibinfo {year} {2011})}\BibitemShut {NoStop}%
\bibitem [{\citenamefont {Al\'en}\ \emph {et~al.}(2003)\citenamefont {Alen},
  \citenamefont {Bickel}, \citenamefont {Karrai}, \citenamefont {Warburton},\
  and\ \citenamefont {Petroff}}]{Alen2003}%
  \BibitemOpen
  \bibfield  {author} {\bibinfo {author} {\bibfnamefont {B.}~\bibnamefont
  {Al\'en}}, \bibinfo {author} {\bibfnamefont {F.}~\bibnamefont {Bickel}},
  \bibinfo {author} {\bibfnamefont {K.}~\bibnamefont {Karrai}}, \bibinfo
  {author} {\bibfnamefont {R.~J.}\ \bibnamefont {Warburton}}, \ and\ \bibinfo
  {author} {\bibfnamefont {P.~M.}\ \bibnamefont {Petroff}},\ }\href {\doibase
  10.1063/1.1609243} {\emph {\bibinfo {title} {Stark-Shift Modulation Absorption Spectroscopy of Single Quantum Dots}}}, {\bibfield  {journal} {\bibinfo  {journal} {Appl.
  Phys. Lett.}\ }\textbf {\bibinfo {volume} {83}},\ \bibinfo {pages} {2235}
  (\bibinfo {year} {2003})}\BibitemShut {NoStop}%
\bibitem [{\citenamefont {Karrai}\ and\ \citenamefont
  {Warburton}(2004)}]{Karrai2003}%
  \BibitemOpen
  \bibfield  {author} {\bibinfo {author} {\bibfnamefont {K.}~\bibnamefont
  {Karrai}}\ and\ \bibinfo {author} {\bibfnamefont {R.~J.}\ \bibnamefont
  {Warburton}},\ }\href {\doibase 10.1016/j.spmi.2004.02.007} {\emph {\bibinfo {title} {Optical Transmission and Reflection Spectroscopy of Single Quantum Dots}}}, {\bibfield
  {journal} {\bibinfo  {journal} {Superlattices and Microstructures}\ }\textbf
  {\bibinfo {volume} {33}},\ \bibinfo {pages} {311} (\bibinfo {year}
  {2003})}\BibitemShut {NoStop}%
\bibitem [{\citenamefont {Acosta}\ \emph {et~al.}(2012)\citenamefont {Acosta},
  \citenamefont {Santori}, \citenamefont {Faraon}, \citenamefont {Huang},
  \citenamefont {Fu}, \citenamefont {Stacey}, \citenamefont {Simpson},
  \citenamefont {Ganesan}, \citenamefont {Tomljenovic-Hanic}, \citenamefont
  {Greentree}, \citenamefont {Prawer},\ and\ \citenamefont
  {Beausoleil}}]{Acosta:2012}%
  \BibitemOpen
  \bibfield  {author} {\bibinfo {author} {\bibfnamefont {V.~M.}\ \bibnamefont
  {Acosta}}, \bibinfo {author} {\bibfnamefont {C.}~\bibnamefont {Santori}},
  \bibinfo {author} {\bibfnamefont {A.}~\bibnamefont {Faraon}}, \bibinfo
  {author} {\bibfnamefont {Z.}~\bibnamefont {Huang}}, \bibinfo {author}
  {\bibfnamefont {K.-M.~C.}\ \bibnamefont {Fu}}, \bibinfo {author}
  {\bibfnamefont {A.}~\bibnamefont {Stacey}}, \bibinfo {author} {\bibfnamefont
  {D.~A.}\ \bibnamefont {Simpson}}, \bibinfo {author} {\bibfnamefont
  {K.}~\bibnamefont {Ganesan}}, \bibinfo {author} {\bibfnamefont
  {S.}~\bibnamefont {Tomljenovic-Hanic}}, \bibinfo {author} {\bibfnamefont
  {A.~D.}\ \bibnamefont {Greentree}}, \bibinfo {author} {\bibfnamefont
  {S.}~\bibnamefont {Prawer}}, \ and\ \bibinfo {author} {\bibfnamefont {R.~G.}\
  \bibnamefont {Beausoleil}},\ }\href {\doibase 10.1103/PhysRevLett.108.206401}
  {\emph {\bibinfo {title} {Dynamic Stabilization of the Optical Resonances of Single Nitrogen-Vacancy Centers in Diamond}}}, {\bibfield  {journal} {\bibinfo  {journal} {Phys. Rev. Lett.}\ }\textbf
  {\bibinfo {volume} {108}},\ \bibinfo {pages} {206401} (\bibinfo {year}
  {2012})}\BibitemShut {NoStop}%
\bibitem [{\citenamefont {Akopian}\ \emph {et~al.}(2013)\citenamefont
  {Akopian}, \citenamefont {Trotta}, \citenamefont {Zallo}, \citenamefont
  {Kumar}, \citenamefont {Atkinson}, \citenamefont {Rastelli}, \citenamefont
  {Schmidt},\ and\ \citenamefont {Zwiller}}]{Akopian:2013}%
  \BibitemOpen
  \bibfield  {author} {\bibinfo {author} {\bibfnamefont {N.}~\bibnamefont
  {Akopian}}, \bibinfo {author} {\bibfnamefont {R.}~\bibnamefont {Trotta}},
  \bibinfo {author} {\bibfnamefont {E.}~\bibnamefont {Zallo}}, \bibinfo
  {author} {\bibfnamefont {S.}~\bibnamefont {Kumar}}, \bibinfo {author}
  {\bibfnamefont {P.}~\bibnamefont {Atkinson}}, \bibinfo {author}
  {\bibfnamefont {A.}~\bibnamefont {Rastelli}}, \bibinfo {author}
  {\bibfnamefont {O.~G.}\ \bibnamefont {Schmidt}}, \ and\ \bibinfo {author}
  {\bibfnamefont {V.}~\bibnamefont {Zwiller}},\ }\href@noop {} {\emph {\bibinfo {title} {An Artificial Atom Locked to Natural Atoms}}}, {\bibfield
  {journal} {\bibinfo  {journal} {arXiv:1302.2005}\ } (\bibinfo {year}
  {2013})}\BibitemShut {NoStop}%
\bibitem [{\citenamefont {Matthiesen}\ \emph {et~al.}(2012)\citenamefont
  {Matthiesen}, \citenamefont {Vamivakas},\ and\ \citenamefont
  {Atat\"ure}}]{Matthiesen2012}%
  \BibitemOpen
  \bibfield  {author} {\bibinfo {author} {\bibfnamefont {C.}~\bibnamefont
  {Matthiesen}}, \bibinfo {author} {\bibfnamefont {A.~N.}\ \bibnamefont
  {Vamivakas}}, \ and\ \bibinfo {author} {\bibfnamefont {M.}~\bibnamefont
  {Atat\"ure}},\ }\href {\doibase 10.1103/PhysRevLett.108.093602} {\emph {\bibinfo {title} {Subnatural Linewidth Single Photons from a Quantum Dot}}}, {\bibfield
  {journal} {\bibinfo  {journal} {Phys. Rev. Lett.}\ }\textbf {\bibinfo
  {volume} {108}},\ \bibinfo {pages} {093602} (\bibinfo {year}
  {2012})}\BibitemShut {NoStop}%
\bibitem [{\citenamefont {Kuhlmann}\ \emph
  {et~al.}(2013{\natexlab{b}})\citenamefont {Kuhlmann}, \citenamefont {Houel},
  , \citenamefont {Brunner}, \citenamefont {Ludwig}, \citenamefont {D.Reuter},
  \citenamefont {Wieck},\ and\ \citenamefont {Warburton}}]{andiRSI}%
  \BibitemOpen
  \bibfield  {author} {\bibinfo {author} {\bibfnamefont {A.~V.}\ \bibnamefont
  {Kuhlmann}}, \bibinfo {author} {\bibfnamefont {J.}~\bibnamefont {Houel}}, ,
  \bibinfo {author} {\bibfnamefont {D.}~\bibnamefont {Brunner}}, \bibinfo
  {author} {\bibfnamefont {A.}~\bibnamefont {Ludwig}}, \bibinfo {author}
  {\bibnamefont {D.Reuter}}, \bibinfo {author} {\bibfnamefont {A.~D.}\
  \bibnamefont {Wieck}}, \ and\ \bibinfo {author} {\bibfnamefont {R.~J.}\
  \bibnamefont {Warburton}},\ }\href@noop {} {\emph {\bibinfo {title} {A Dark-Field Microscope for Background-Free Detection of Resonance Fluorescence from Single Semiconductor Quantum Dots Operating in a Set-And-Forget Mode}}}, {\bibfield  {journal} {\bibinfo
  {journal} {arXiv:1303.2055}\ } (\bibinfo {year}
  {2013}{\natexlab{b}})}\BibitemShut {NoStop}%
\bibitem [{\citenamefont {Warburton}\ \emph {et~al.}(2000)\citenamefont
  {Warburton}, \citenamefont {Sch\"aflein}, \citenamefont {Haft}, \citenamefont
  {Bickel}, \citenamefont {Lorke}, \citenamefont {Karrai}, \citenamefont
  {Garcia}, \citenamefont {Schoenfeld},\ and\ \citenamefont
  {Petroff}}]{Warburton2000}%
  \BibitemOpen
  \bibfield  {author} {\bibinfo {author} {\bibfnamefont {R.~J.}\ \bibnamefont
  {Warburton}}, \bibinfo {author} {\bibfnamefont {C.}~\bibnamefont
  {Sch\"aflein}}, \bibinfo {author} {\bibfnamefont {D.}~\bibnamefont {Haft}},
  \bibinfo {author} {\bibfnamefont {F.}~\bibnamefont {Bickel}}, \bibinfo
  {author} {\bibfnamefont {A.}~\bibnamefont {Lorke}}, \bibinfo {author}
  {\bibfnamefont {K.}~\bibnamefont {Karrai}}, \bibinfo {author} {\bibfnamefont
  {J.~M.}\ \bibnamefont {Garcia}}, \bibinfo {author} {\bibfnamefont
  {W.}~\bibnamefont {Schoenfeld}}, \ and\ \bibinfo {author} {\bibfnamefont
  {P.~M.}\ \bibnamefont {Petroff}},\ }\href {\doibase 10.1038/35016030}{\emph {\bibinfo {title} {Optical Emission from a Charge-Tunable Quantum Ring}}},
  {\bibfield  {journal} {\bibinfo  {journal} {Nature (London)}\ }\textbf {\bibinfo
  {volume} {405}},\ \bibinfo {pages} {926} (\bibinfo {year}
  {2000})}\BibitemShut {NoStop}%
\bibitem [{\citenamefont {Latta}\ \emph {et~al.}(2009)\citenamefont {Latta},
  \citenamefont {H\"ogele}, \citenamefont {Zhao}, \citenamefont {Vamivakas},
  \citenamefont {Maletinsky}, \citenamefont {Kroner}, \citenamefont {Dreiser},
  \citenamefont {Carusotto}, \citenamefont {Badolato}, \citenamefont {Schuh},
  \citenamefont {Wegscheider}, \citenamefont {Atature},\ and\ \citenamefont
  {Imamoglu}}]{Latta2009}%
  \BibitemOpen
  \bibfield  {author} {\bibinfo {author} {\bibfnamefont {C.}~\bibnamefont
  {Latta}}, \bibinfo {author} {\bibfnamefont {A.}~\bibnamefont {H\"ogele}},
  \bibinfo {author} {\bibfnamefont {Y.}~\bibnamefont {Zhao}}, \bibinfo {author}
  {\bibfnamefont {A.~N.}\ \bibnamefont {Vamivakas}}, \bibinfo {author}
  {\bibfnamefont {P.}~\bibnamefont {Maletinsky}}, \bibinfo {author}
  {\bibfnamefont {M.}~\bibnamefont {Kroner}}, \bibinfo {author} {\bibfnamefont
  {J.}~\bibnamefont {Dreiser}}, \bibinfo {author} {\bibfnamefont
  {I.}~\bibnamefont {Carusotto}}, \bibinfo {author} {\bibfnamefont
  {A.}~\bibnamefont {Badolato}}, \bibinfo {author} {\bibfnamefont
  {D.}~\bibnamefont {Schuh}}, \bibinfo {author} {\bibfnamefont
  {W.}~\bibnamefont {Wegscheider}}, \bibinfo {author} {\bibfnamefont
  {M.}~\bibnamefont {Atature}}, \ and\ \bibinfo {author} {\bibfnamefont
  {A.}~\bibnamefont {Imamoglu}},\ }\href {\doibase 10.1038/NPHYS1363}{\emph {\bibinfo {title} {Confluence of Resonant Laser Excitation and Bidirectional Quantum-Dot Nuclear-Spin Polarization}}},
  {\bibfield  {journal} {\bibinfo  {journal} {Nat. Phys.}\ }\textbf
  {\bibinfo {volume} {5}},\ \bibinfo {pages} {758} (\bibinfo {year}
  {2009})}\BibitemShut {NoStop}%
\bibitem [{\citenamefont {Nagourney}(2010)}]{nag}%
  \BibitemOpen
  \bibfield  {author} {\bibinfo {author} {\bibfnamefont {W.}~\bibnamefont
  {Nagourney}},\ }\href@noop {} {\emph {\bibinfo {title} {Quantum Electronics
  for Atomic Physics}}}\ (\bibinfo  {publisher} {Oxford University Press},\
  \bibinfo {year} {2010})\BibitemShut {NoStop}%
\bibitem [{\citenamefont {Kogan}(1996)}]{Kogan}%
  \BibitemOpen
  \bibfield  {author} {\bibinfo {author} {\bibfnamefont {S.}~\bibnamefont
  {Kogan}},\ }\href@noop {} {\emph {\bibinfo {title} {Electronic Noise and
  Fluctuations in Solids}}}\ (\bibinfo  {publisher} {Cambridge University
  Press, London},\ \bibinfo {year} {1996})\BibitemShut {NoStop}%
\bibitem [{\citenamefont {Al\'en}\ \emph {et~al.}(2006)\citenamefont {Alen},
  \citenamefont {H\"ogele}, \citenamefont {Kroner}, \citenamefont {Seidl},
  \citenamefont {Karrai}, \citenamefont {Warburton}, \citenamefont {Badolato},
  \citenamefont {Medeiros-Ribeiro},\ and\ \citenamefont {Petroff}}]{Alen2006}%
  \BibitemOpen
  \bibfield  {author} {\bibinfo {author} {\bibfnamefont {B.}~\bibnamefont
  {Al\'en}}, \bibinfo {author} {\bibfnamefont {A.}~\bibnamefont {H\"ogele}},
  \bibinfo {author} {\bibfnamefont {M.}~\bibnamefont {Kroner}}, \bibinfo
  {author} {\bibfnamefont {S.}~\bibnamefont {Seidl}}, \bibinfo {author}
  {\bibfnamefont {K.}~\bibnamefont {Karrai}}, \bibinfo {author} {\bibfnamefont
  {R.~J.}\ \bibnamefont {Warburton}}, \bibinfo {author} {\bibfnamefont
  {A.}~\bibnamefont {Badolato}}, \bibinfo {author} {\bibfnamefont
  {G.}~\bibnamefont {Medeiros-Ribeiro}}, \ and\ \bibinfo {author}
  {\bibfnamefont {P.~M.}\ \bibnamefont {Petroff}},\ }\href {\doibase
  10.1063/1.2354431} {\emph {\bibinfo {title} {Absorptive and Dispersive Optical Responses of Excitons in a Single Quantum Dot}}}, {\bibfield  {journal} {\bibinfo  {journal} {Appl. Phys.
  Lett,}\ }\textbf {\bibinfo {volume} {89}},\ \bibinfo {pages} {123124}
  (\bibinfo {year} {2006})}\BibitemShut {NoStop}%
\bibitem [{\citenamefont {Atat\"ure}\ \emph {et~al.}(2007)\citenamefont
  {Atat\"ure}, \citenamefont {Dreiser}, \citenamefont {Badolato},\ and\
  \citenamefont {Imamoglu}}]{Atatuere2007}%
  \BibitemOpen
  \bibfield  {author} {\bibinfo {author} {\bibfnamefont {M.}~\bibnamefont
  {Atat\"ure}}, \bibinfo {author} {\bibfnamefont {J.}~\bibnamefont {Dreiser}},
  \bibinfo {author} {\bibfnamefont {A.}~\bibnamefont {Badolato}}, \ and\
  \bibinfo {author} {\bibfnamefont {A.}~\bibnamefont {Imamoglu}},\ }\href
  {\doibase 10.1038/nphys521} {\emph {\bibinfo {title} {Observation of Faraday Rotation from a Single Confined Spin}}}, {\bibfield  {journal} {\bibinfo  {journal}
  {Nat. Phys.}\ }\textbf {\bibinfo {volume} {3}},\ \bibinfo {pages} {101}
  (\bibinfo {year} {2007})}\BibitemShut {NoStop}%
\end{thebibliography}

%

\end{document}